\documentclass[prb,superscriptaddress,twocolumn,floatfix,amsmath,amssymb,showpacs]{revtex4-1}

\usepackage{graphicx}
\usepackage{dcolumn}
\usepackage{bm}
\usepackage{color}
\hfuzz=\maxdimen
\begin{document}

\title{Weak metal-metal transition in the vanadium oxytelluride Rb$_{1-\delta}$V$_2$Te$_2$O}


\author{Abduweli Ablimit}
\affiliation{Department of Physics, Zhejiang University, Hangzhou 310027, China}
\author{Yun-Lei Sun}
\affiliation{School of Information and Electrical Engineering, Zhejiang University City College, Hangzhou 310015, China}
\affiliation{State Key Lab of Silicon Materials, Zhejiang University, Hangzhou 310027, China}
\author{Hao Jiang}
\affiliation{School of Physics and Optoelectronics, Xiangtan University, Xiangtan 411105, China}
\author{Si-Qi Wu}
\affiliation{Department of Physics, Zhejiang University, Hangzhou 310027, China}
\author{Ya-Bin Liu}
\affiliation{Department of Physics, Zhejiang University, Hangzhou 310027, China}
\author{Guang-Han Cao}
\email[corresponding author: ]{ghcao@zju.edu.cn}
\affiliation{Department of Physics, Zhejiang University, Hangzhou 310027, China}
\affiliation{State Key Lab of Silicon Materials, Zhejiang University, Hangzhou 310027, China}
\affiliation{Collaborative Innovation Centre of Advanced Microstructures, Nanjing University, Nanjing 210093, China}

\date{\today}

\begin{abstract}
We report the synthesis, crystal structure, physical properties, and first-principles calculations of a vanadium-based oxytelluride Rb$_{1-\delta}$V$_2$Te$_2$O ($\delta\approx0.2$). The crystal structure bears two-dimensional V$_2$O square nets sandwiched with tellurium layers, mimicking the structural motifs of cuprate and iron-based superconductors. The material exhibits metallic conductivity with dominant hole-type charge carriers. A weak metal-to-metal transition takes place at $\sim$100 K, which is conformably characterized by a slight kink/hump in the electrical resistivity, jumps in the Hall and Seebeck coefficients, a minute drop in the magnetic susceptibility, and a small peak in the heat capacity. Neither Bragg-peak splitting nor superlattice reflections can be detected within the resolution of conventional x-ray diffractions. The band-structure calculations show that V-3$d$ orbitals dominate the electronic states at around Fermi energy where a $d_{yz}/d_{xz}$ orbital polarization shows up. There are three Fermi-surface sheets that seem unfavorable for nesting. Our results suggest an orbital or spin-density-wave order for the low-temperature state and, upon suppression of the competing order, emergence of superconductivity could be expected.
\end{abstract}



\maketitle
\section{\label{sec:level1}Introduction}
Cuprate high-temperature superconductors are structurally featured with two-dimensional CuO$_2$ planes which serve as the key motif for superconductivity. For this reason, great efforts have been devoted to exploring superconducting materials in layered 3$d$-transition-metal compounds. The most prominent advance of the exploration is the discovery of iron-based superconductors which contain planar Fe-square nets that are sandwiched by pnictogen or chalcogen (hereafter denoted by $X$) monolayers. In these fascinating materials, superconductivity emerges as the competing orders, such as antiferromagnetism and charge- or spin-density wave (DW), are sufficiently suppressed~\cite{rmp_Scalapino,Gabovich}.

Along this line, materials containing $M_2$O square planes, where $M$ stands for $3d$-transition elements such as Ti~\cite{ZAAC,BaTi2As2O.cxh,2223Ti.cxh,Ba22241,CsTi2Te2O}, Fe~\cite{2223Fe.Mayer,22221Fe,2221Fe.CGF,1221Fe.HP,2323.cxl}, Co~\cite{2223Co.wc}, Mn~\cite{2223Mn} and V~\cite{CsV2S2O}, have attracted much interest~\cite{rev_stam,review.Chu,review.Yajima}. The $M_2$O planes, being of an anti-CuO$_2$-type (the cations and anions exchange their sites) structure, are sandwiched by $X$ layers, forming a unique structural motif $M_2$O$X_2$ with mixed anions. The $M_2$O$X_2$ block layers thus capture the structural characteristics of both copper- and iron-based superconductors. Those compounds with $M$ = Fe, Co and Mn are always antiferromagnetic insulators~\cite{2223Fe.fmh,22221Fe,2221Fe.CGF,1221Fe.HP,2323.cxl,2223Co.wc,2223Mn}, in which the $M$ ion tends to have a high-spin state~\cite{2223Fe.neutron,22221Fe,1221Fe.HP,2223Co.neutron,2223Mn} that seems unfavorable for superconductivity. Materials with $M$ = Ti mostly show a metallic behavior, and undergo a possible DW transition~\cite{rev_stam,review.Chu,review.Yajima}. There have been a lot of debates on the nature of the DW transition~\cite{2221.1997,2221Ti.neutron,Na2Ti2Sb2O.cxh,2221.lzy,2221.NMR.ljl,2221.fdl,OO.Kim,1221.Singh,1221.wgt,Ivanovskii,1221.mSR,1221.fdl}. The most recent results confirm a charge-density-wave (CDW)~\cite{2221.Raman,2221cdw.Davies,2221.DFT2017} or a nematic charge order~\cite{1221.nc,1221.bond-order}, depending on specific material system.

BaTi$_2$As$_2$O exhibits a metal-to-metal transition (MMT) at 200 K~\cite{BaTi2As2O.cxh}, accompanying with a nematic charge ordering~\cite{1221.nc}. As arsenic is replaced with antimony, the transition temperature is lowered to $\sim$50 K~\cite{BaTi2Sb2O.Yajima,BaNaTi2Sb2O} and, remarkably, the long-sought superconductivity emerges at $T_{\mathrm{c}}=1.2$ K in BaTi$_2$Sb$_2$O~\cite{BaTi2Sb2O.Yajima}. The $T_{\mathrm{c}}$ value can be increased appreciably up to 6.1 and 4.6 K, respectively, by a hole doping~\cite{BaNaTi2Sb2O,BaK1221,BaRb1221,1221Sn} and by applying a ``negative chemical pressure" through an isovalent Bi-for-Sb substitution~\cite{BaTi2Pn2O,BaTi2SbBiO}. Notably, superconductivity was \emph{exclusively} observed in the 1221-type Ti-based oxypnictides~\cite{Ti2Bi2O}.

Isostructural to BaTi$_{2}X_2$O ($X=$ As, Sb, Bi), CsV$_2$S$_2$O was recently reported as the first vanadium-based member in the 1221-structure family~\cite{CsV2S2O}. Unlike BaTi$_{2}X_2$O series, the compound is intrinsically ``doped", as the formal valence state of vanadium is +2.5. The material shows a bad-metal behavior with a resistivity minimum at around 177 K, and no bulk superconductivity was observed above 1.8 K. Motivated by the negative chemical-pressure effect that enhances superconductivity in the Ti-based materials~\cite{BaTi2Sb2O.Yajima,BaTi2SbBiO,BaTi2Pn2O}, we attempted to prepare an isostructural V-based oxytelluride in order to explore possible superconductivity. As a result, Rb$_{1-\delta}$V$_2$Te$_2$O ($\delta\approx0.2$) with a nominal vanadium valence of +2.6 was successfully synthesized. Unexpectedly, the material exhibits a MMT at $T^{\ast}\sim100$ K, evidenced from the measurements of transport, magnetic, and thermodynamic properties. Meanwhile, no structural phase transition was detected within the measurement resolution of conventional x-ray diffractions. The first-principles calculations were also carried out, supplying useful information on the nature of the MMT.

\section{\label{sec:level2}Methods}
\emph{Sample's synthesis.} Rb$_{1-\delta}$V$_2$Te$_2$O polycrystals were synthesized via a solid state reaction using source materials of Rb (99.5\%), V$_2$O$_5$ (99.99\%), Te (99.999\%) and V (99.5\%). The stoichiometric (i.e., in the ideal formula RbV$_2$Te$_2$O) mixture was placed into an alumina tube, the latter of which was sealed in an evacuated quartz ampule. The sample-loaded ampule was slowly heated to 300 $^{\circ}$C, holding for 10 hours. After the first-stage reaction, the mixture was homogenized and pressed into pellets, and then were loaded in an alumina tube again. Next, the sample-loaded alumina tube was jacketed with a Ta tube that was sealed under argon atmosphere. The Ta tube was then sealed in an evacuated quartz ampule. The sample assembly was sintered at 720 $^{\circ}$C for 33 hours in a muffle furnace, followed by switching off the furnace. We found that the sample's quality could be further improved by repeating the sintering process. Note that the final products were non-stoichiometric with Rb deficiency and, the Rb deficiency varied a little (within 10\%) for each synthesis. Also noted is that the product is very air sensitive, and all the operations of sample handling were carried out in a glove box with water and oxygen content less than 1 ppm.

\emph{Crystal structure determination.} The as-prepared sample was structurally characterized by powder X-ray diffractions (XRD) at temperatures from 300 K down to 20 K, carried out on a PANalytical diffractometer (Empyrean Series 2) with a monochromatic Cu-$K_{\alpha1}$ radiation. To avoid the sample's deterioration, we mixed the sample with $N$-type grease in the Ar-filled glove box prior to the XRD measurement. The crystal-structure parameters of Rb$_{1-\delta}$V$_2$Te$_2$O were refined by a Rietveld analysis using the code RIETAN-2000~\cite{rietan}. We tried different profile functions, and the best one was the modified split pseudo-Voigt function~\cite{rietan}. Note that the weighted profile R-factor, $R_{\mathrm{wp}}$, is relatively large, presumably because of the disturbance of the grease mixed. The chemical composition of the crystallites in samples was checked using an energy-dispersive x-ray (EDX) spectrometer (Model Octane Plus) equipped in a field-emission scanning electron microscope (Hitachi S-4800).

\emph{Physical property measurements.} The electrical-transport and heat-capacity measurements were performed on a Physical Property Measurement System (PPMS-9). In the transport measurements, the sample pellet was cut into a thin rectangular bar, on which electrodes were made using silver paste. A standard four-electrode method was employed for the resistivity measurement. For the Hall measurement, a cross-like four-wire configuration (van der Pauw technique), was adopted. The Hall coefficient was obtained by permutating the voltage and current terminals~\cite{Hall_measurement} under a magnetic field of 60 kOe. The Seebeck coefficient was measured by a conventional steady-state technique with a temperature gradient of $\sim$1 K/cm. The magnetization was measured on a Magnetic Property Measurement System (MPMS-5) under a magnetic field  of 10 kOe.

\emph{Electronic structure calculations.} The density functional theory based first-principles calculations were conducted employing the Vienna Ab-initio Simulation Package (VASP)~\cite{VASP}. The exchange correlations were treated with the generalized gradient approximation (GGA)~\cite{GGA}. We found that the crystal-structure relaxation well reproduces the experimental parameters, and the latter (those of Rb$_{0.83}$V$_2$Te$_2$O at 300 K) were adopted for the calculation. The plane-wave basis energy cutoff was set at 550 eV. For the band structure and density-of-states (DOS) calculations, we generated $\mathbf{k}$ meshes of $15\times 15\times 7$ and $18\times18 \times 6$, respectively, using the scheme of Monkhorst and Pack~\cite{BZ}.

\section{\label{sec:level3}Results}

\subsection{\label{subsec:level1}Synthesis and crystal structure}

\begin{figure}
\includegraphics[width=8cm]{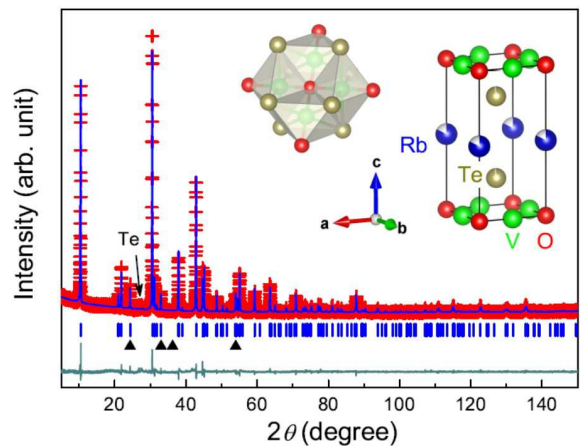}
\caption{Rietveld refinement profile of the powder x-ray diffraction at room temperature for a Rb$_{1-\delta}$V$_2$Te$_2$O ($\delta=$ 0.17) sample. The main Bragg reflections of the possible V$_2$O$_3$ impurity are marked with solid triangles. The central inset shows the V$_2$Te$_2$O layers made of face-sharing VTe$_4$O$_2$ polyhedra, and the left inset displays a unit cell of the crystal structure.}
\label{xrd}
\end{figure}

We prepared the target material of RbV$_2$Te$_2$O many times. All the as-prepared samples contain dominantly the 1221-type phase. The secondary phases such as Te and V$_2$O$_3$ were typically below 5\%, as estimated from the XRD reflection intensities. Meanwhile, the lattice constants vary a little ($\sim$0.3\%) from sample to sample in different batches. Analyses of the EDX spectra of the samples' crystallites confirm the existence of Rb, V, Te and O in the sample. The resultant molar ratios of the constituent heavy elements (oxygen is not included here because of the relatively large measurement uncertainty) were Rb : V : Te = 0.40(5) : 1.0 : 0.92(5), indicating that there is substantial Rb deficiency and the Te deficiency is subtle. It is the uncontrollable slight difference in Rb deficiency that leads to the variation in lattice constants. Note that our attempt to synthesize stoichiometric RbV$_2$Te$_2$O using extra amount of Rb was not successful. The Rb-deficient Rb$_{1-\delta}$V$_2$Te$_2$O is likely to be the thermodynamically stable phase under the synthesis condition. This situation resembles the case in Cs$_{1-x}$Ti$_2$Te$_2$O with $x\approx0.2$~\cite{CsTi2Te2O}. In this paper, we report the experimental data on two of the Rb$_{1-\delta}$V$_2$Te$_2$O samples with $\delta=$ 0.17 and 0.22.

\begin{table}
\caption{Crystallographic data of Rb$_{0.83}$V$_2$Te$_2$O obtained from the structural refinement ($R_{\mathrm{wp}}$ = 11.98\%, $R_{\mathrm{B}}$ = 6.07\%, $R_{\mathrm{F}}$ = 4.52\%, and $S$ = 1.48)~\cite{rietan} of powder x-ray diffraction at room temperature.}
  \label{tab1}\renewcommand\arraystretch{1.3}
  \begin{tabular}{ccccccc}
      \hline \hline
    \multicolumn{3}{c}{Chemical Formula} & &  & \multicolumn{2}{c}{Rb$_{0.83}$V$_2$Te$_{1.94}$O}\\
\multicolumn{3}{c}{Space Group} & & & \multicolumn{2}{c}{$P$4/$mmm$ (No. 123)} \\
\multicolumn{3}{c}{$a$ (\r{A})} & & & \multicolumn{2}{c}{4.0524(1)} \\
\multicolumn{3}{c}{$c$ (\r{A})} & & & \multicolumn{2}{c}{8.4334(2)} \\
    \hline
    Atoms  & site  & $x$ & $y$ & $z$ & Occupancy& $U_{\mathrm{iso}}$ ({\AA}$^2$) \\
    Rb       & 1b  & 0  & 0  & 1/2  & 0.83(1) & 0.0178(9) \\
    V        & 2f  & 0  & 1/2  & 0 & 1  & 0.0155(9) \\
    Te       & 2h  & 1/2  & 1/2  & 0.2288(1) & 0.97(1)  & 0.0134(4)\\
    O        & 1a  & 0 & 0 & 0 & 1 & 0.019(4) \\
    \hline
    \hline
  \end{tabular}
\end{table}

Figure~\ref{xrd} shows the XRD pattern of a typical Rb$_{1-\delta}$V$_2$Te$_2$O sample. The XRD profile can be well reproduced by the Rietveld refinement. The crystallographic data refined are presented in Table~\ref{tab1}, in which the occupancy factors of Rb and Te are given to be 0.83(1) and 0.97(1), respectively, consistent with the EDX measurement (hereafter we will not consider the slight Te vacancy). The room-temperature lattice constants of the other sample Rb$_{0.78}$V$_2$Te$_2$O are, $a$ = 4.0414(2) \AA~ and $c$ = 8.4610(5) \AA, which are about 0.3\% smaller and larger, respectively. The result seems reasonable because the decrease in Rb content practically increases the formal valence of vanadium, the latter of which should shorten the bond distances primarily in the basal plane.

Table~\ref{tab2} lists the crystal-structure parameters of Rb$_{0.83}$V$_2$Te$_2$O, in comparison with those of other 1221-type titanium/vanadium oxypnictides/oxychalcogenides. The cell constants are larger than the counterparts of CsV$_2$S$_2$O, yet smaller than those of Cs$_{0.8}$Ti$_2$Te$_2$O, which can be basically understood in terms of ionic radii~\cite{radius}. Nevertheless, the axial ratio $c/a$ increases remarkably upon the isovalent substitutions with heavier elements at the pnictogen/chalcogen site (e.g., Te substitutes for S), in agreement with the result in the solid-solution system BaTi$_2$(Sb$_{1-x}$Bi$_x$)$_2$O~\cite{BaTi2SbBiO}. The anisotropic crystal-structure deformation by the isovalent substitutions is related to the highly anisotropic chemical bonding, according to the first-principles calculations~\cite{Ivanovskii}. The chemical bonds within $M_2$O$X_2$ block layers are mixtures of metallic, covalent, and ionic contributions, whilst the interlayer bonding is basically ionic. Note that the $z$ coordinate of chalcogens of V-based compounds is particularly small, compared with those of Ti-based materials. This directly leads to an obviously longer $X-X$ interatomic distance along $c$ direction, suggesting enhanced two dimensionality in the V-based oxychalcogenides.

\begin{table*}
\caption{Comparison of structural and physical properties of titanium/vanadium ($M$) oxypnictides/oxychalcogenides with BaTi$_2$As$_2$O-type structure. $X$ stands for a pnictogen or chalcogen, whose fractional coordinate is (1/2, 1/2, $z_X$). $n_{3d}$ is the $3d$-electron counts, obtained by assuming the valence states of Ba$^{2+}$, $A^{+}$ ($A$ = Cs and Rb), $Pn^{3-}$ ($Pn$ = As, Sb, and Bi), and $Ch^{2-}$ ($Ch$ = S and Te). $T^{\ast}$ and $T_{\mathrm{sc}}$ denote the density-wave-like and superconducting transition temperatures, respectively.   Some related interatomic distances and angle are given. $X$-$X$ is the interatomic distance along $c$ direction. $X$-$M$-$X$ is the bond angle along $a$ or $b$ direction.}
  \label{tab2}\renewcommand\arraystretch{1.3}
  \begin{tabular}{lccccccccccr}
      \hline\hline
      Compounds& $a$ ({\AA}) & $c$ ({\AA})& $z_X$& $c/a$ &$M$-$M$ ({\AA})&$X$-$X$ ({\AA})&$X$-$M$-$X$ ($^{\circ}$)& $n_{3d}$& $T^{\ast}$ (K) & $T_{\mathrm{sc}}$ (K) &Refs. \\
       \hline
    BaTi$_2$As$_2$O & 4.047(3) & 7.275(5)&0.2440(1) & 1.798 &2.862&3.725&97.5 &1&200 & - &  \cite{BaTi2As2O.cxh} \\
    BaTi$_2$Sb$_2$O & 4.11039(2)  & 8.08640(4)&0.24779(5) & 1.967  &2.906&4.079& 91.5&1  &50 & 1.2  &\cite{BaTi2Sb2O.Yajima} \\
    BaTi$_2$Bi$_2$O & 4.1236(4)  & 8.3447(1) &0.2513(2) & 2.024 & 2.916&4.151& 89.0&1 &- & 4.6  &  \cite{BaTi2Pn2O}\\
    Cs$_{0.8}$Ti$_2$Te$_2$O & 4.0934(3) & 8.9504(9)& 0.2776(5)& 2.187 &2.894&3.981& 79.0&1.4  &- & - &\cite{CsTi2Te2O} \\
    \hline
     CsV$_2$S$_2$O & 3.9455(1) & 7.4785(1)&0.2117(2) & 1.895&2.790&4.312&102.5 &2.5 & - & - & \cite{CsV2S2O} \\
     Rb$_{0.83}$V$_2$Te$_2$O & 4.0524(1) & 8.4334(2) &0.2288(1)& 2.081 &2.866&4.574&92.8 &2.42 & 100 & - & This work\\
\hline\hline
  \end{tabular}
\end{table*}


Like the case in CsV$_2$S$_2$O~\cite{CsV2S2O}, vanadium is coordinated by two O$^{2-}$ and four Te$^{2-}$ ions, forming a \emph{trans}-octahedron in Rb$_{1-\delta}$V$_2$Te$_2$O. The Te$-$V$-$Te bond angles are nearly 90$^{\circ}$, and therefore, the VTe$_4$O$_2$ polyhedron looks like a uniaxially compressed regular octahedron. The interatomic distances of V$-$O and V$-$Te are 2.026 \AA~ and 2.798 \AA, respectively. Each VTe$_4$O$_2$ octahedron shares two of its faces, taking the shape of two-dimensional layers (see the inset in Fig.~\ref{xrd}). This geometrical configuration gives rise to a relatively short V$-$V distance (2.866 \AA), suggesting existence of V$-$V metallic bonding.

\subsection{\label{subsec:level2}Physical properties}

\begin{figure}
\includegraphics{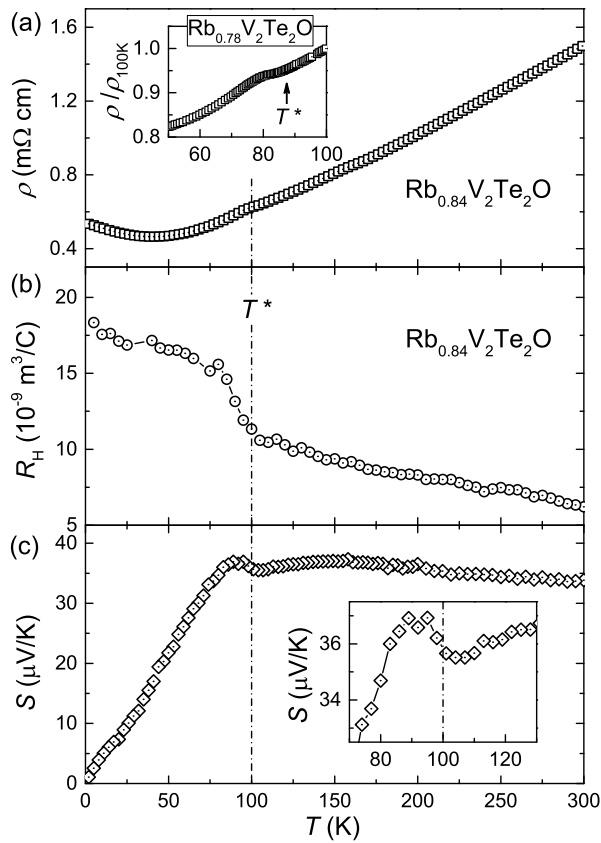}
\caption{Temperature dependence of electrical resistivity $\rho$ (a), Hall coefficient $R_\mathrm{H}$ (b), and thermoelectric power $S$ (c) for the Rb$_{0.83}$V$_2$Te$_2$O polycrystalline sample. The dot dash line marks the onset of the anomalies. The inset of (a) shows the resistivity anomaly at a lower $T^{\ast}$ in Rb$_{0.78}$V$_2$Te$_2$O.}
\label{transport}
\end{figure}

Figure~\ref{transport}(a) shows the temperature dependence of resistivity for the as-prepared Rb$_{1-\delta}$V$_2$Te$_2$O polycrystalline samples. The material behaves as a metal in the sense of the positive temperature coefficient of $\rho(T)$ above 40 K. A slight resistivity upturn appears below 40 K, the reason of which is unknown. As a whole, the resistivity value is around $\sim$1 m$\Omega$ cm (the absolute resistivity could be somewhat lower for the single-crystal sample), suggesting that the material could be a bad metal. There is a detectable kink at $T^{*}\sim$100 K, similar to, yet much less pronounced than, those of the spin-density-wave (SDW) transition in parent compounds of iron-based superconductors such as BaFe$_2$As$_2$~\cite{Ba122.cxh}. For samples of Rb$_{1-\delta}$V$_2$Te$_2$O with lower Rb content, the resistivity kink evolves into a small hump at a lowered temperature [see the inset of Fig.~\ref{transport}(a)], resembling those of the MMT in BaTi$_2$As$_2$O~\cite{BaTi2As2O.cxh} and BaTi$_2$Sb$_2$O~\cite{BaTi2Sb2O.Yajima,BaNaTi2Sb2O}. Note that such an anomaly is absent in the vanadium oxysulfide CsV$_2$S$_2$O~\cite{CsV2S2O}.

The temperature dependence of Hall coefficient ($R_\mathrm{H}$) for the same Rb$_{0.83}$V$_2$Te$_2$O sample is shown in Fig.~\ref{transport}(b). The $R_\mathrm{H}$ value is positive in the whole temperature range, indicating dominant hole-type conduction. This result contrasts with the electron-type dominated transport in the Ti-based compounds such as Na$_2$Ti$_2$$X_2$O ($X$ = As, P)~\cite{Na2Ti2Sb2O.cxh}. Further, if assuming a single-band scenario that satisfies the simple relation, $n_\mathrm{H}=1/(eR_\mathrm{H})$, one obtains a ``net" hole concentration of $n_\mathrm{H}\sim6\times10^{26}$ m$^{-3}$ at 100 K. The $n_\mathrm{H}$ value is equivalent to a Hall number of 0.04 per vanadium atom, which is much reduced compared with the $3d$-electron counts (see Table~\ref{tab2}). The result thus strongly suggests a multiband scenario. Coincident with the resistivity kink at the $T^{*}$, a significant rise in the $R_\mathrm{H}(T)$ data shows up, suggesting a partial gap opening on the Fermi surface (FS). The gap opening ordinarily leads to a resistivity jump or hump because of the decrease in carrier concentration. Here the resistivity kink observed can be explained in terms of an enhanced carrier mobility for the low-temperature state, similar to the case in the parent compounds of iron-based superconductors~\cite{122SDW.wnl}.

The apparent hole-type transport is further confirmed by the positive values of Seebeck coefficient ($S$) in the whole temperature range, as shown in Fig.~\ref{transport}(c). At low temperatures, the $S(T)$ relation is almost linear with a slope of d$S$/d$T$ $\approx$ 0.45 $\mu$V K$^{-2}$. The slope is an order of magnitude larger than that of an ordinary metal (e.g., d$S$/d$T$ = 0.024 $\mu$V K$^{-2}$ calculated with the single-band formula $S=\frac{\pi^{2}k_{\mathrm{B}}^{2}}{3e}\frac{T}{\epsilon_{\mathrm{F}}}$ for a typical Fermi energy of $\epsilon_{\mathrm{F}}\sim$ 1 eV~\cite{Behnia}). This discrepancy suggests again a multiband scenario which allows an effectively small $\epsilon_{\mathrm{F}}$. The high-temperature $S$ values are of the order of $k_\mathrm{B}/e$ (86 $\mu$V K$^{-1}$), suggestive of a bad-metal behavior again. In the inset of Fig.~\ref{transport}(c), a jump of $S(T)$ is clearly seen at the $T^{\ast}$. A similar anomaly in $S(T)$ was also observed at the MMT in the parent compounds of iron-based superconductors~\cite{1111Co.wc} and in the Ti-based compounds such as Na$_2$Ti$_2$Sb$_2$O~\cite{Na2Ti2Sb2O.cxh}.

\begin{figure}
\includegraphics{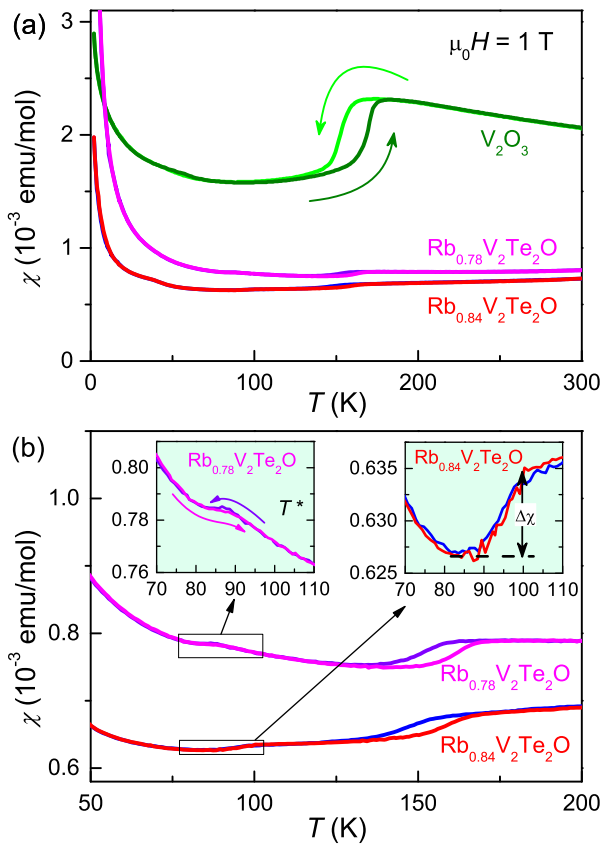}
\caption{Temperature dependence of magnetic susceptibility for the reference material V$_2$O$_3$ (as purchased) as well as two samples of Rb$_{1-\delta}$V$_2$Te$_2$O ($\delta=$ 0.17 and 0.22). Panel (b) highlights the susceptibility anomalies, with insets showing the close-ups of the anomalies at 90 and 100 K, respectively.}
\label{MT}
\end{figure}

Figure~\ref{MT} shows the temperature dependence of magnetic susceptibility for Rb$_{1-\delta}$V$_2$Te$_2$O ($\delta=$ 0.17 and 0.22). In the high-temperature regime, the $\chi(T)$ data are basically temperature independent, characteristic of Pauli paramagnetism (rather than Curie-Weiss paramagnetism). The apparent upward tails at low temperatures, which also appear in the reference sample of V$_2$O$_3$, are probably due to small amount of paramagnetic impurities. On closer examination, as explicitly shown in Fig.~\ref{MT}(b), one sees two steps at around $\sim$100 and $\sim$150 K, respectively. Since there is no obvious anomaly in the transport properties and in the heat capacity (see below) around 150 K, we suspect that the 150 K anomaly may come from the secondary phase V$_2$O$_3$, the latter of which undergoes a first-order antiferromagnetic transition just at $\sim$150 K~\cite{V2O3}. The molar fraction of V$_2$O$_3$ impurity is estimated to be $\sim$4\% (from the relative magnitude of the susceptibility drop), which basically agrees with the XRD result.

The second anomaly appears at $T^{\ast}\approx$ 90 K and 100 K, respectively, for the two Rb$_{1-\delta}$V$_2$Te$_2$O samples with $\delta=$ 0.22 and 0.17. The susceptibility drop ($\Delta \chi$) suggests a DW-like gap opening at the FS. The magnitude of $\Delta \chi$ is only about 8$\times10^{-6}$ emu/mol, about an order of magnitude lower than the counterparts (6 - 20$\times10^{-5}$ emu/mol) of the Ti-based compounds~\cite{BaTi2As2O.cxh,BaTi2Sb2O.Yajima,BaNaTi2Sb2O,Na2Ti2Sb2O.cxh}. This smaller value of $\Delta \chi$ is actually consistent with the less pronounced anomalies in resistivity, Hall coefficient and Seebeck coefficient, all of which point to a small fraction of gap opening at the FS. Here we note that the transition seems to be of weak first order because of the thermal hysteresis with $\Delta T\approx$ 2 K.

\begin{figure}
\includegraphics{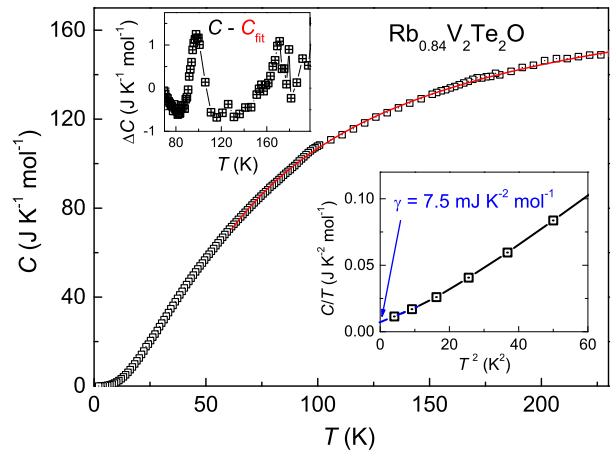}
\caption{Temperature dependence of specific heat for Rb$_{0.83}$V$_2$Te$_2$O. The red solid line is a polynomial fit, with which  the specific-heat difference, $\Delta C=C-C_{\mathrm{fit}}$, is obtained as displayed in the upper inset. The lower inset plots $C/T$ versus $T^2$ in the low-temperature region.}
\label{sh}
\end{figure}

Figure~\ref{sh} shows the temperature dependence of specific heat for Rb$_{0.83}$V$_2$Te$_2$O. A weak anomaly at around 100 K is discernable from the raw $C(T)$ data. The anomaly turns out to be an obvious peak (see the upper inset), after the ``background" data (represented by a polynomial fit) are subtracted out. The $C(T)$ peak also features a weak first-order transition. The relatively low value of the $C(T)$ peak indicates a small latent heat, which accounts for the small thermal hysteresis observed.

The lower inset of Fig.~\ref{sh} plots $C/T$ versus $T^2$ in the low-temperature region. The deviation from the expected linearity is possibly due to additional specific-heat contributions such as Schottky anomalies. Nevertheless, one may estimate the Sommerfeld coefficient as $\gamma_{\mathrm{exp}}\approx$ 7.5 mJ K$^{-2}$ mol$^{-1}$, using a linear extrapolation. This non-zero $\gamma$ value indicates a metallic ground state for the low-temperature phase in Rb$_{0.83}$V$_2$Te$_2$O, albeit of the resistivity upturn below 40 K.

\subsection{\label{subsec:level3}Low temperature XRD}

\begin{figure}
\includegraphics{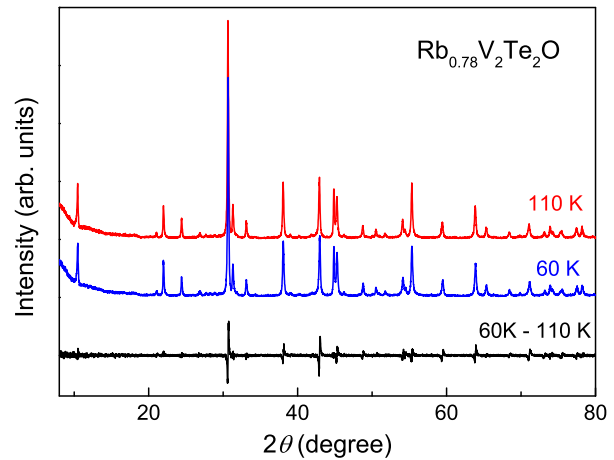}
\caption{Powder x-ray diffraction patterns at 110 and 60 K for Rb$_{0.78}$V$_2$Te$_2$O. The bottom curve is their difference from which no new reflections can be detected within the instrumental sensitivity.}
\label{ltxrd1}
\end{figure}

\begin{figure*}
\includegraphics[width=15cm]{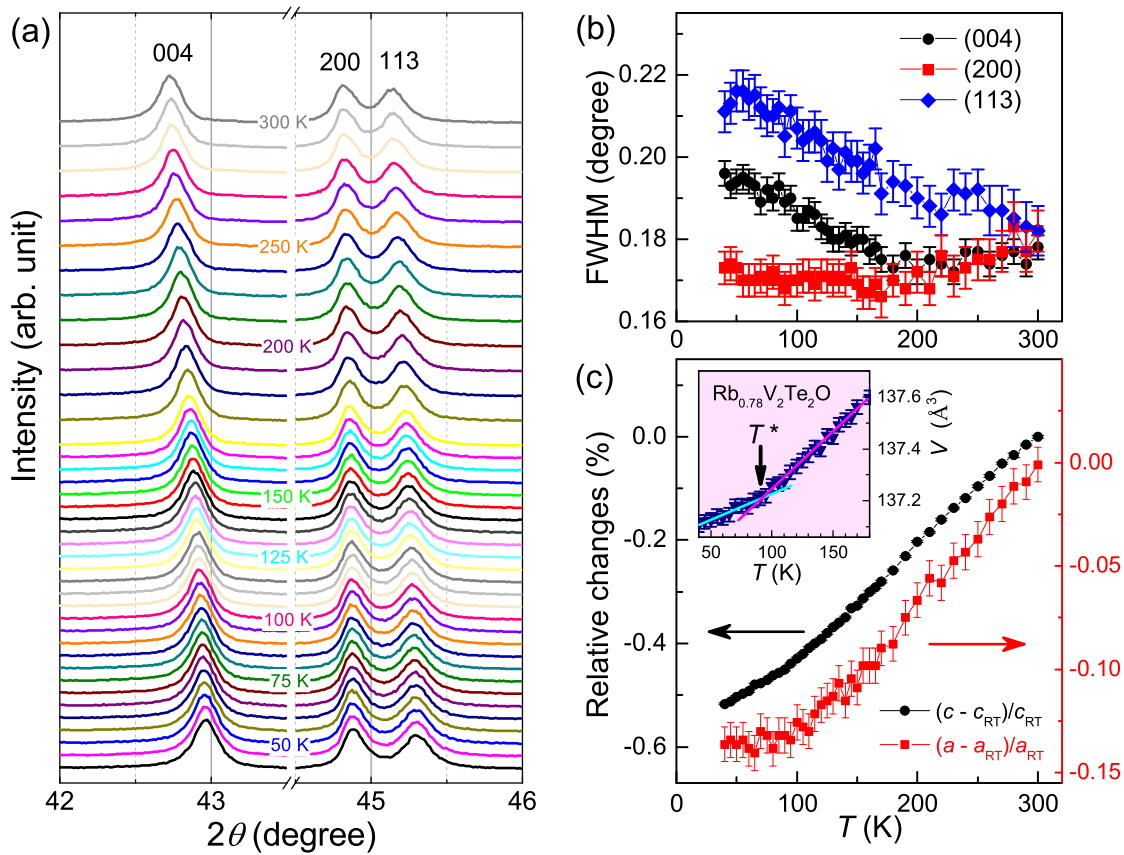}
\caption{Low-temperature X-ray diffractions for Rb$_{0.78}$V$_2$Te$_2$O. (a) Evolution of the (004), (200) and (113) Bragg reflections with decreasing temperature. (b) Temperature dependence of FWHMs (Full Width at Half Maximum). (c) Relative changes in lattice parameters $a$ (right axis) and $c$ (left axis) with respect to those at room temperature. The inset plots the unit-cell volume as a function of temperature.}
\label{ltxrd2}
\end{figure*}

We conducted low-temperature XRD measurements to investigate a possible structural transition at the $T^{\ast}$. Fig.~\ref{ltxrd1} shows the XRD patterns of Rb$_{0.78}$V$_2$Te$_2$O at 110 and 60 K, both of which look essentially the same (except for a slight shift due to thermal expansion of lattice). To detect any possible new reflections, we also plot their difference (bottom line). No additional Bragg peaks are discernable, suggesting absence of any superstructure for the low-temperature phase within the detection limit ($\sim$1\% of the intensity of the strongest reflection).

Then, is there any tetragonal-to-orthorhombic structural transition at the $T^{\ast}$, like the case in BaTi$_2$As$_2$O~\cite{1221.nc}? The symmetry-broken lattice distortion commonly leads to a peak splitting~\cite{1111.lyk,1221.nc}. So, we focus on the XRD profiles in the $2\theta$ range that contains (200) and (113) reflections. Fig.~\ref{ltxrd2}(a) shows evolution of the XRD profiles with decreasing temperature for Rb$_{0.78}$V$_2$Te$_2$O. No peak splitting can be directly seen. We further investigated the possible variation in peak width, conventionally coined as Full Width at Half Maximum (FWHM). As shown in Fig.~\ref{ltxrd2}(b), upon cooling down through $T^{\ast}\approx$ 90 K, there is no sudden rise in FWHM. From the uncertainty of the extracted FWHMs, the assumed tetragonal-orthorhombic distortion, ($a-b$)/$a$, if exists, would be less than 0.04\%. This means that the speculated lattice distortion is so small as being even comparable to the uncertainty of lattice parameters, suggesting possible absence of any lattice symmetry breaking.

Let us examine the temperature dependence of lattice constants. Here the lattice constants $c$ and $a$ were derived, respectively, on the basis of the exact positions of (004) and (200) peaks, the latter of which were obtained by the data fitting using pseudo-Voigt function. Fig.~\ref{ltxrd2}(c) shows the relative changes in $a$ and $c$ (with respect to those at 300 K). While both parameters decrease with temperature, the decrease in $c$ axis is over three times faster, indicating a highly anisotropic thermal expansion. The result is consistent with relatively weak chemical bonding along the $c$ axis.

Notably, both $a$ and $c$ exhibit a discernable kink at around the $T^{\ast}$. The kink in $a$ also reflects the change in V$-$V interatomic distance, since $d_{\mathrm{V}-\mathrm{V}}=\sqrt{2}a/2$. This anomaly is evident in the temperature dependence of unit-cell volume [see the inset of Fig.~\ref{ltxrd2}(c)]. The volumetric thermal expansion coefficient, defined by $\alpha_V=(\partial V/\partial T)_P/V$, changes from 32 ppm K$^{-1}$ ($T\rightarrow T^{\ast}_+$) to 15 ppm K$^{-1}$ ($T\rightarrow T^{\ast}_-$). Now that the MMT is probably associated with a partial gapping of the Fermi surface, i.e., a change in the electronic states, the remarkable reduction in $\alpha_V$ at the MMT suggests a substantial coupling of electronic instability with lattice.

\subsection{Electronic structure calculation}

The possible magnetically ordered ground state was first investigated in our first-principle calculations. The result of GGA calculation with spin-orbit coupling suggests a G-type (an antiparallel alignment of magnetic moments between nearest neighbors) antiferromagnetic ground state with a magnetic moment of 1.65 $\mu_\mathrm{B}$/V-atom along the crystallographic $a$ or $b$ axis. The calculated magnetic moment is much larger than that of BaTi$_{2}$Sb$_2$O (0.2 $\mu_\mathrm{B}$/Ti-atom)~\cite{1221.Singh,1221.wgt}, yet significantly lower than the fully localized high-spin value of 2.5 $\mu_\mathrm{B}$/V-atom, suggesting a SDW order with considerable itinerant $3d$ electrons. Note that this G-type antiferromagnetic SDW remains when the on-site Coulomb interactions are included (GGA$+U$). Nevertheless, such a SDW ordering would give rise to a large magnetic susceptibility drop at $T^{\ast}$, which contradicts with the experimental observation. In the present circumstance that lacks a clear experimental conclusion on the magnetic ground state, here we only present the results for the nonmagnetic state.

\begin{figure}
\includegraphics[width=8cm]{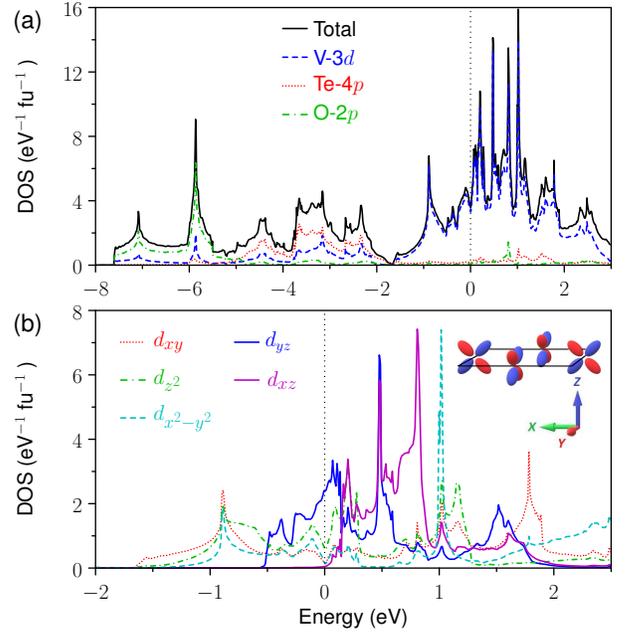}
\caption{Energy dependence of electronic density of states (DOS) that are projected onto the related atoms (a) and V-$3d$ orbitals (b) for nonmagnetic RbV$_2$Te$_2$O. The $d$-orbital projection is based on the crystalline-lattice coordinate (i.e., $x\leftrightarrow a$, $y\leftrightarrow b$, and $z\leftrightarrow c$). The electronic state at the Fermi energy shows $d_{yz}$/$d_{xz}$ orbital polarization, as depicted in the upper-right corner of the bottom panel.}
\label{dos}
\end{figure}

Fig.~\ref{dos}(a) shows the electronic DOS projected onto the related atoms for the stoichiometric RbV$_2$Te$_2$O. First, the oxygen-$2p$ and tellurium-$4p$ states are mostly filled, consistent with the formal valence states of O$^{2-}$ and Te$^{2-}$. The O-$2p$ and Te-$4p$ states are located at around 6 and 3.5 eV below $E_\mathrm{F}$, respectively. These valence states are appreciably hybridized with the vanadium-$3d$ orbitals. By contrast, the DOS around $E_\mathrm{F}$ (say, $|E-E_\mathrm{F}|\leq 1$ eV) are contributed almost from the V-$3d$ states only. This means that the conduction bands are basically composed of the V-$3d$ metallic-bonding and non-bonding states. Similar phenomena are seen in the iron-based superconductor system~\cite{1111.Singh}. However, the pnictogen $p$ states in BaTi$_{2}$Sb$_2$O contribute a relatively large DOS at $E_\mathrm{F}$~\cite{1221.Singh,1221.wgt,Ivanovskii}.

The DOS at $E_{\mathrm{F}}$, $D(E_{\mathrm{F}})$, is 3.58 eV$^{-1}$fu$^{-1}$, which corresponds to a bare electronic specific-heat coefficient of, $\gamma_{0}=\frac{1}{3}\pi^{2}k_{\mathrm{B}}^2D(E_{\mathrm{F}})=8.4$ mJ K$^{-2}$ mol$^{-1}$ (the $D(E_{\mathrm{F}})$ value for the real composition of Rb$_{0.8}$V$_2$Te$_2$O is expected to increase a little in the rigid-band scenario). The calculated $\gamma_{0}$ is very close to the experimental value $\gamma_{\mathrm{exp}}$ ($\sim$7.5 mJ K$^{-2}$ mol$^{-1}$), suggesting insignificant enhancement of effective mass of carriers. Besides, the bare Pauli magnetic susceptibility,  $\chi_0=\mu_\mathrm{B}^{2}D(E_{\mathrm{F}})=1.2\times10^{-4}$ emu mol$^{-1}$, is an order of magnitude higher than the magnetic susceptibility drop $\Delta\chi$ at $T^{\ast}$, further indicating that the energy gap opens only on a small part of the FS.

Now that the V-$3d$ orbitals dominantly contribute the electronic states around $E_\mathrm{F}$, we further look into the electron-filling distribution in each of the $3d$ orbitals. As is shown in Fig.~\ref{dos}(b), the $d_{xy}$, $d_{z^{2}}$ and $d_{x^{2}-y^{2}}$ states are all partially filled, with a V-shape minimum in DOS at $E_\mathrm{F}$. The $d_{yz}$ orbital is the dominant contributor at $E_\mathrm{F}$. In contrast, there is scarcely electron occupation in the $d_{xz}$ orbital. This result strongly suggests a $d_{yz}/d_{xz}$ orbital polarization at $E_\mathrm{F}$. The orbital polarization is characterized by the occupation (unoccupation) of electrons in V(1/2, 0, 0) $d_{yz}$ ($d_{xz}$) and  V(0, 1/2, 0) $d_{xz}$ ($d_{yz}$)orbitals. Notably, this global orbital polarization will not break the tetragonal symmetry, consistent with the experimental observation.

\begin{figure}
\includegraphics[width=8cm]{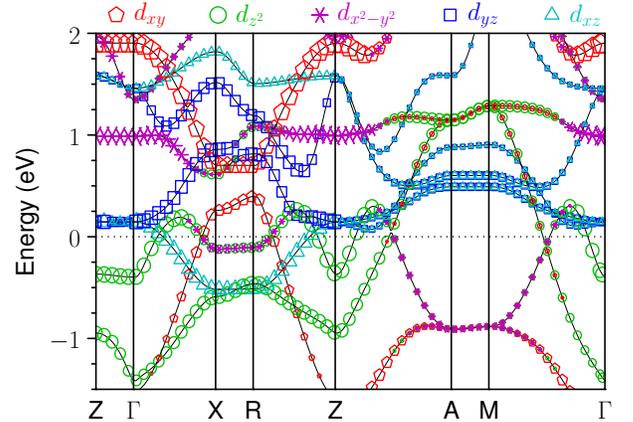}
\caption{Electronic band structure projected onto different V-$3d$ orbitals for RbV$_2$Te$_2$O.}
\label{band}
\end{figure}

\begin{figure}
\includegraphics[width=8cm]{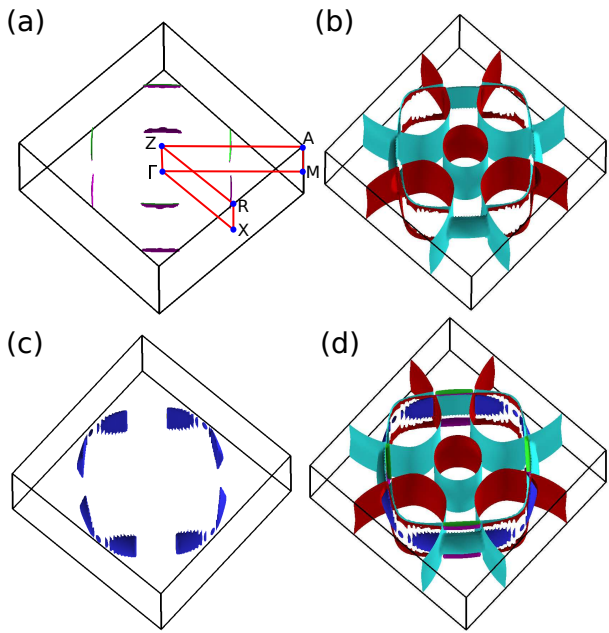}
\caption{Three-dimensional views of the calculated Fermi-surface (FS) sheets of $\alpha$ (a), $\beta$ (b), and $\gamma$ (c) bands for nonmagnetic RbV$_2$Te$_2$O. The merged Fermi surface is shown in (d).}
\label{fs}
\end{figure}

The calculated band structure is shown in Fig.~\ref{band}, which is also projected on different V-$3d$ orbitals. Along the horizontal directions of $\Gamma$X, ZR, ZA, and $\Gamma$M in the Brillouin zone [see Fig.~\ref{fs}(a)], the bands cross the $E_\mathrm{F}$ level. In contrast, no band crossing happens and, the energy dispersion is very weak nearby the $E_\mathrm{F}$, along the vertical directions of $\Gamma$Z, XR and MA, indicating two dimensionality in the electronic structure. The band around $\Gamma$Z line is electron-like. Indeed, one of the FS sheets ($\beta$ band) shows a electron-type cylinder centered at $\Gamma$ point [Fig.~\ref{fs}(b)]. However, there are large portions with hole-like dispersion in the $\Gamma$M and ZA lines, in accordance with the holelike occupancy in the $\beta$ FS sheet mainly along the diagonal direction of the Brillouin Zone. There are band inversions with linear dispersions close to $E_\mathrm{F}$ at several $\mathbf{k}$ points. In the $\alpha$ band, for example, hole-type dispersion can be seen at round the midpoint of ZA line, which gives rise to the small hole pockets  [Fig.~\ref{fs}(a)].

As shown in Fig.~\ref{fs}, the FS of nonmagnetic RbV$_2$Te$_2$O comprises of three sheets. The largest FS sheet from the $\beta$ band is very two dimensional with a complex topology. The volumes of electron-type and hole-type pockets are comparable, which explains the small Hall number measured experimentally. The hole-type $\alpha$ and electron-type $\gamma$ sheets turn out to be Fermi patches with very small volumes. The overall FS, shown in Fig.~\ref{fs}(d), is basically unfavorable for nesting, which could be the reason for the absence of superlattice in Rb$_{1-\delta}$V$_2$Te$_2$O.

\section{\label{sec:level4}Discussion and conclusion}

The physical property measurements above indicate that Rb$_{1-\delta}$V$_2$Te$_2$O undergoes a weak MMT at $T^{\ast}\approx$ 100 K. The MMT is accompanied with a slight resistivity kink or hump, jumps in Hall and Seebeck coefficients, a small drop in magnetic susceptibility, and a small peak in heat capacity. These electrical and magnetic anomalies point to a partial gap opening over the FS. Nevertheless, no evidence of structural phase transition is detectable in the conventional low-temperature XRD experiments. At the same time, the thermal expansion coefficient shows a discontinuity at the $T^{\ast}$.

The MMT in Rb$_{1-\delta}$V$_2$Te$_2$O bears both similarities and differences with those of the analogous Ti-based oxypnictides~\cite{rev_stam,review.Chu,review.Yajima} and the iridium ditelluride IrTe$_2$~\cite{Matsumoto1999}. The non-doped Ti-based oxypnictides show similar, yet much enhanced, anomalies in electrical transport, magnetic susceptibility, and specific heat~\cite{BaTi2As2O.cxh,BaTi2Sb2O.Yajima,BaNaTi2Sb2O,Na2Ti2Sb2O.cxh}. Partial gap opening on the Fermi surface or Fermi patches was observed from angle-resolved photoemission spectroscopy (ARPES)~\cite{2221.fdl,2221cdw.Davies,1221.fdl}. Recent investigations have found different types of charge orders. For Na$_2$Ti$_2$Sb$_2$O and Na$_2$Ti$_2$As$_2$O, superstructures with the propagation vectors of (1/2, 0, 0) and (0, 1/2, 1/2) were revealed respectively by single-crystal x-ray diffractions, corroborating the CDW ground state. As for 1221-type BaTi$_2$As$_2$O and BaTi$_2$Sb$_2$O, an accompanying CDW superstructure is absent. Instead, tetragonal-symmetry breaking was clearly evidenced from neutron diffraction, which was explained in terms of an intra-unit-cell nematic charge order~\cite{1221.nc}. The MMT in IrTe$_2$ is also accompanied with an obvious structural transformation~\cite{Matsumoto1999} in which Ir dimerizations were revealed~\cite{Ir-dimer}. Here in Rb$_{1-\delta}$V$_2$Te$_2$O, however, neither superlattice nor crystal-symmetry breaking is observed. One possibility is that the structural modification is so slight (with relative change less than 0.04\%) that cannot be detected by conventional XRD instrument. If so, the energy gain related to the slight lattice distortion will be very tiny accordingly (like the case in Jahn-Teller distortion). This tiny energy gain might not be able to account for the transition temperature as high as 100 K.

Notably, the $3d$-electron count in Rb$_{1-\delta}$V$_2$Te$_2$O is nearly 2.5, which is 1.5 electrons more than those of the non-doped Ti-based oxypnictides. This leads to a remarkable difference in the state occupancy on each $3d$ orbital. In fact, the band structure and DOS of RbV$_2$Te$_2$O and BaTi$_2$Sb$_2$O~\cite{1221.Singh,1221.wgt} would be similar, if the Fermi level of latter were lifted upward by $\sim$0.7 eV. In BaTi$_2$Sb$_2$O, both $d_{yz}$ and $d_{xz}$ orbitals contribute little to the DOS at $E_\mathrm{F}$~\cite{1221.wgt}, namely, there is no $d_{yz}/d_{xz}$ orbital polarization. In the case of RbV$_2$Te$_2$O, $d_{yz}$ contributes the most (72\%), while $d_{xz}$ contributes the least (1.6\%), primarily because of the electron filling. This could gives rise to an orbital polarization without global crystal-symmetry breaking. Note that this kind of orbital polarization may be driven by spin-orbit coupling as well as Coulomb interactions~\cite{OO.Kim}.

As a general trend, V-$3d$ electrons bear stronger Coulomb interactions than Ti-$3d$ electrons do. Thus, the vanadium atom in Rb$_{1-\delta}$V$_2$Te$_2$O could carry a considerably large moment (as our calculation shows), and the material might be magnetically ordered in its ground state. If this is the case, the most probable magnetic order is the G-type antiferromagnetic SDW, in which the crystal-symmetry also remains. Future studies with various techniques such as neutron diffraction, high-resolution synchrotron XRD, nuclear magnetic resonance, and muon spin relaxation are expectable to clarify the nature of the weak MMT.

In summary, we have successfully synthesized a layered vanadium oxytelluride Rb$_{1-\delta}$V$_2$Te$_2$O ($\delta\approx$ 0.2). The compound crystallizes in a BaTi$_2$As$_2$O-type structure containing V$_2$O square nets. The structural and physical-property measurements indicate a weak first-order MMT with no expected crystal-symmetry breaking observed. The MMT could be either a SDW ordering or an orbital ordering/polarization, as suggested from the experimental measurements and electronic-structure calculations. Notably, Rb$_{1-\delta}$V$_2$Te$_2$O shows similarities and/or relevances in crystal-structure, electronic-structure, and physical properties with the known cuprate~\cite{rmp_Scalapino}, iron-based~\cite{rmp_Scalapino}, Ti-based~\cite{BaTi2Sb2O.Yajima,BaNaTi2Sb2O}, and Ir-based~\cite{Ir-SC1,Ir-SC2,Ir-SC3} superconductors. In this context, it is hopeful to explore superconductivity via suppression of the MMT in the present system.

\begin{acknowledgments}
This work was supported by the National Key Research and Development Program of China (No. 2017YFA0303002), the State Key Lab of Silicon Materials, China (SKL2017-12), and the Fundamental Research Funds for the Central Universities of China.
\end{acknowledgments}


%

\end{document}